\documentclass[12pt,preprint]{aastex}

\usepackage{amsfonts}

\shorttitle{The Origin of Tycho's SN}
\shortauthors{F.J. Lu et al.}

\begin{document}

\title{The Single-degenerate Binary Origin of Tycho's Supernova as Traced by the Stripped Envelope of the Companion}

\author{F.J. Lu\altaffilmark{1}, Q.D. Wang\altaffilmark{2}, M.Y. Ge\altaffilmark{1},
J.L. Qu\altaffilmark{1}, X.J. Yang\altaffilmark{3}, S.J. Zheng\altaffilmark{1}, and Y. Chen\altaffilmark{1}}

\affil{$^1$Key Laboratory for Particle Astrophysics, Institute of High Energy
Physics, Chinese Academy of Sciences, Beijing 100049, P.R. China;
lufj@mail.ihep.ac.cn}
\affil{$^2$Department of Astronomy, University of Massachusetts, Amherst, MA 01003, USA}
\affil{$^3$Faculty of Materials, Optoelectronics and Physics, Xiangtan University, Hunan 411105, P.R. China}

%\affil{$^2$National Astronomical
%Observatories, Chinese Academy of Sciences, Beijing 100012, P.R.
%China}
%
\begin{abstract}
We propose that a non-thermal X-ray arc inside the remnant of
Tycho's supernova (SN) represents the interaction between the SN ejecta and
the companion star's envelope lost in the impact of the explosion. The X-ray emission of the
remnant further shows an apparent shadow casted by the arc in the
opposite direction of the explosion site, consistent with the
blocking of the SN ejecta by the envelope. This scenario supports
the single degenerate binary origin of Tycho's SN. The properties of
the X-ray arc, together with the previous detection of the companion
candidate and its space velocity by Ruiz-Lapuente et al. (2004) and
Hern\'{a}ndez et al.(2009), enables us to further infer 1) the
progenitor binary has a period of 4.9$^{+5.3}_{-3.0}$ days, 2) the
companion gained a kick velocity of 42$\pm$30 km s$^{-1}$, and 3) the
stripped envelope mass is about 0.0016($\leq0.0083$) $M_{\sun}$. However,
we notice that the nature of the companion candidate is still 
under debate, and the above parameters need to be revised according to
the actual properties of the companion candidate. Further work to measure the 
proper motion of the arc and to check the capability of the interaction to emit 
the amount of X-rays observed from the arc is also needed to validate the current scenario. 
\end{abstract}


\section{Introduction}
While Type Ia supernovae (Ia SNe) play a fundamental role in
cosmology and chemical evolution of the Universe, their exact origin
remains greatly uncertain. One of the leading mechanisms for such a
SN is the thermonuclear explosion of a white dwarf when its mass
reaches a critical value via accretion from a normal stellar
companion in a binary, which is called the single-degenerate
scenario \citep{Whelan1973,Wheeler1975,Nomoto1982,Hillebrandt2000}.
This scenario is supported by the possible identification of the
survived stellar companion of the Ia SN observed by Tycho Brahe in
1572 \citep{Ruiz2004}, although little is yet known about the
putative progenitor binary. A specific prediction of the
single-degenerate scenario is that up to 0.5 solar mass can be
stripped by the impact of Ia SN from its companion star
\citep{Wheeler1975,Fryxell1981,Taam1984,Chugaj1986,Marietta2000,Meng2007}.
So far, however, no direct observational evidence for the stripped
envelope has been reported; only an upper limit of 0.01 solar mass
has been set for two extragalactic Ia SNe \citep{Leonard2007}. In
comparison, stellar remnants are expected to be completely dispersed
if a SN is due to the merger of two white dwarfs (the
double-degenerate scenario; Iben \& Tutukov  1984; Whelan \& Iben
1973).

  We select the Tycho supernova remnant (SNR) to search for the stripped
material motivated by the detection of its candidate companion star
(Tycho G, a G0-G2 type subgiant), which shows a large peculiar
velocity \citep{Ruiz2004,Hern2009}. This remnant is one of the two
identified Galactic type Ia SNRs, as revealed by the light curve,
radio emission, and X-ray spectra
\citep{Baade1945,Baldwin1957,Hughes1995,Ruiz2004}. It is young (437
years), nearby (3$\pm$1 kpc; de Vaucouleurs 1985), and of high X-ray
brightness (Cassam-Chena\"{i} et al. 2007). With the available deep
observations from the Chandra X-ray Observatory (CXO), this remnant is an
ideal laboratory to search for the stripped mass entrained in the
ejecta.

\section{Observation and Data Reduction}

Table 1 lists the 12 CXO observations of the
Tycho SNR used in this study. These observations were carried out by
the ACIS-I, the imagine array of the Advanced CCD Imaging Spectrometer
(ACIS), with a field of view large enough to cover the whole
remnant. The data were calibrated with the Chandra Interactive
Analysis of Observations (CIAO V4.1) software package following the
standard procedure to correct for charge-transfer inefficiency (CTI)
effects and the time-dependence of the gain, to clean bad pixels,
and to remove time intervals of background flares. The final total
effective exposure for these observations is a little bit longer
than 1 Ms.

\section{Results}

Fig. 1 shows the intensity images of the remnant in different energy
bands. In the 4-6 keV band image (Fig. 1 (a)), which is dominated by
non-thermal X-ray emission sensitive to shocks, which accelerate
particles, there appears an intriguing arc (as marked in the
figure), only about half way from the SN site (RA (2000)
=00:25:23.8; DEC (2000) =64:08:04.7; Ruiz-Lapuente et al. 2004a) and is as
bright, narrow, and sharp as those filaments at the outer boundaries
of the remnant. This unusual arc was first noticed by Warren et al.
(2005). They suggested that it may still be part of the SNR rim seen
in projection. In comparison, Figs. 1 (b), (c), and (d) show the remnant
in 1.6-2.0, 2.2-2.6, and 6.2-6.8 keV energy bands, representing the
intensity distributions of Si, S, and Fe emission lines. We find that the
Fe K$\alpha$ line emission  is unusually faint in a cone
just outside of the arc (in the direction away from the SN site) , in
comparison with the other regions beyond the same radius.
The opening angle of this cone (about 20$^{\circ}$) is similar to that
of the arc relative to the SN site.
In the same cone, especially at the outer rims of the SNR, the 4-6 keV
continuum emission as well as the Si and S line intensities also appear
to be relatively deficient, though not as obvious as in the Fe
K$\alpha$ band. These relative intensity contrasts are shown
quantitatively in Fig. 2 (a). On the other hand, Fig 2(b) shows the same
intensity contrasts for region within the arc radius, where local peaks
appear at nearly the same angular positions of the dips of the profiles
outside the arc. Observing the images we find that the enhancements are
very close to the arc.
We interpret the intensity deficiency
in the cone outside the arc and the local enhancements inside the arc
as the blocking of the SN ejecta by the arc.

The spectrum of the arc is shown in Fig 3. The on-source spectrum is
extracted from the region defined by the inner polygon in Fig. 1(a),
while the background is from the region between the inner and outer
polygons.   Fig 3 shows obvious dips at the positions of the Si, S
and Fe lines in the source spectrum obtained. Since the background thermal emission
dominates the nonthermal emission at these energies, these dips are probably
due to the clumpy distribution of the SN ejecta and so
highly variable background thermal emission surrounding the arc.
%This means that the background spectrum changes significantly across
%the remnant and the arc region is relatively deficient of line
%emission (and perhaps thermal continuum emission) compared to the
%regions surrounding the arc.
Fitting the source spectrum with a power law model gives a photon index
of 2.45$\pm$0.09 (90\% confidence errors), an absorption column
density of (8.7$\pm$0.9)$\times$10$^{21}$ cm$^{-2}$, an unabsorbed
0.5-10 keV flux of 7.8$\times$10$^{-13}$ erg cm$^{-2}$ s$^{-1}$, and
$\chi^2$ of 291 for 200 degrees of freedom. If we ignore the data in
1.6-2.0, 2.2-2.6 and 6.0-7.0 keV, which are probably effected by the
over-subtraction of the background line emission, the fitted
parameters are then photon index 2.47$^{+0.16}_{-0.08}$, absorption
column density 9.0$^{+1.4}_{-0.7}\times$10$^{21}$ cm$^{-2}$,
and unabsorbed 0.5-10 keV flux 8.4$\times$10$^{-13}$ erg
cm$^{-2}$~s$^{-1}$, as well as a significantly improved $\chi^2$ of 176 for
171 degrees of freedom. Therefore, we conclude that the arc is
nonthermal.

\section{The nature and origin of the X-ray arc}
We find that the X-ray arc is most probably in the interior of the
SNR instead of a (morphologically) unusual feature of the outer rim
projected well inside the remnant.  The sharp and bright appearance
of the X-ray arc is similar to those of the filaments at the outer
rim of the SNR (Cassam-Chena\"{i} et al. 2007), implying that the arc is
observed almost edge on. If it is in the out layer of the SNR and
the observed small angular distance from the geometric centre is due
to the projection effect, then the arc should be much more diffuse
as it is observed half face on.  In the 4-6 keV map, there exist
some relatively bright features within the blast wave boundary.
However, most of them are substantially more diffuse and coincide
with bright thermal structures spatially, and none of them is as far
away from the boundary as the arc is (see also Warren et al. 2005).
The arc is convexed toward the SN site, which is opposite to those
of the filaments at the outer boundary. Furthermore, larger and fainter
filaments tend to run northward from the southeast (Warren et al. 2005)
and bent outward. The morphology of the arc can be naturally
produced  by the interaction between the SN ejecta and a cloud. As
will be discussed in the following, the arc most probably represents
the materials stripped from the companion star by the SN explosion
and is in the interior of the SNR.

First, the arc must be related to the progenitor system of the SN.
If it is not related to the progenitor system, it should then
represent a dense cold molecular cloud that has survived for long time.
Actually, the milli-meter and optical observations suggest that the Tycho SNR is possibly
interacting with molecular clouds at the northeastern and the
southwestern rims (Lee et al. 2004; Ghavamian et al. 2000). These
regions show strong 4-6 keV emission. However, neither CO nor
optical emission enhancement has been detected to be spatially
coincident with the X-ray arc. It is unlikely to be a molecular
cloud.

Second, the arc cannot arise from the materials ejected by the
progenitor binary system of the SN. Since the mass donor is
suggested to be very similar to the Sun but
 a slightly evolved one (Ruiz-Lapuente et al. 2004), we don't expect that it could
contribute to the cloud (e.g., via a stellar wind). One might think
that a planetary nebula surrounding the
 exploded white dwarf could be a source of the cloud material. However,
planetary nebulae always show a spherically or axially symmetric
morphology.  The singleness of the X-ray arc makes this possibility
very unlikely.

Finally, and most probably, the matter generating the X-ray arc is
stripped from the companion star during the SN explosion. Such a
mechanism has been suggested by many theoretical works (Wheeler et
al. 1975; Fryxell \& Arnett 1981; Taam \& Fryxell 1984; Chugai 1986;
Marietta et al. 2000; Meng et al. 2007) although no direct
observational evidence is currently available (Leonard 2004).
Arguments for this mechanism are as follows: (1) The stripped
materials are confined in a small angular range, which can naturally
explain the singleness of the arc; (2) The opening angle (about
20$\degr$) of the arc relative to the SN site (Ruiz-Lapuente et al.
2004), the absence of the X-ray line emission in the cone away
from the arc, and the local enhancements of X-ray emission immediately 
within the arc radius are all well consistent with the ejecta blocking scenario
of the stripped stellar envelope (Marietta et al. 2000); (3) The gas
in the envelope of the companion star is expected to have a temperature of
a few $10^3$ K, it can not cool down to the CO emitting temperature in 400 years, 
especially in a hot environment; (4) The impact of the SN blast wave and ejecta on the stripped 
envelope will generate a shock wave, just like those represented by the nonthermal X-ray 
filaments at the outer rim, but with a smaller 
velocity because the envelope is denser than the surrounding ISM. The shock wave
can accelerate electrons to emit the nonthermal X-rays. (5) The angle
between the direction of the arc to the explosive centre and the
proper motion velocity of Tycho G is well consistent with the
theoretical predictions and simulations (Marietta et al. 2000;
Ruiz-Lapuente et al. 2004; Meng et al. 2007), as detailed in the
following.

\section{Constraints on the progenitor binary system}
The above interpretation together with the existing measurement of
the companion's present velocity gives tight constraints on
properties of the progenitor binary of Tycho SN (see Fig. 4 for an
illustration). Because of the SN impact, the companion star should
receive a kick as well as the envelope stripping in the same
direction and in the orbital plane of the progenitor binary. The
direction should be perpendicular to the orbiting velocity of the
companion as it is expected to be in a circular orbit just before
the SN \citep{Marietta2000,Meng2007}. For Tycho G, the measured
radial velocity is -50$\pm$10 km s$^{-1}$ [with the projected
Galactic rotation contribution (-30$\pm10$ km s$^{-1}$) subtracted; Hern\'andez et al. 2009],
while the tangential one is -94$\pm27$ km s$^{-1}$ \citep{Ruiz2004}.
Taking the uncertainty in the distance (3$\pm$1 kpc; de Vaucouleurs
1985) into account, the tangential velocity is -94$\pm41$ km
s$^{-1}$. The projected angle between the stripping direction and
the current velocity of Tycho G relative to the SN site is
$\alpha$=63$\pm$13$\degr$, where the error accounts for the
uncertainties in both the proper motion measurement and the
arc/shadow central line (about 2$\degr$). Because the X-ray
arc/shadow is viewed almost edge on (with an assumed uncertainty of
10$\degr$), the stripped velocity should be nearly perpendicular to
the line of sight. Therefore, the real angle between the stripping
(or the kick) and the velocity of Tycho G is $\beta$
=67$\pm$16$\degr$. Numerical simulation by Marietta et al. (2000) shows
that the ratio between the orbital velocity and kick velocity is
from 2.3 to 11.2, and so the real angle 
between the kick and the space velocity of the stellar remnant is between 67$\degr$ to 85$\degr$.
Our result is consistent with these predictions.

We can then infer that the companion had an
orbital velocity ($V_o$) of 98$\pm$36 km s$^{-1}$ in the progenitor
binary, received a kick velocity ($v_k$) of 42$\pm$30 km
s$^{-1}$, and the inclination angle of the orbital plane is
31$\pm13\degr$. We can also constrain the separation between the two
stars in the progenitor binary with the formula
$a=\frac{M_{1}^{2}G}{(M_{1}+M_{2})V_{o}^{2}}$, where $M_{1}$ is the
mass of the SN progenitor and is assumed to be 1.4 solar mass ---
the Chandrasekhar mass of a white dwarf, $M_{2}$ the mass of the
companion star, and $G$ the gravitational constant. Tycho G is similar to
the Sun spectroscopically \citep{Ruiz2004,Hern2009} and thus has a
mass of about one solar or slightly higher (depending on the
luminosity type; see below). We then estimate
$a=\frac{1}{2.4+\varepsilon}(2.7\pm2.0)\times10^{7}$ km, where
$\varepsilon$ is the mass stripped from the companion and should be
considerably smaller than one, hence negligible. The corresponding
orbital period is then 4.9$^{+5.3}_{-3.0}$ days. These orbital
parameters and kick velocity are summarized in Table 2. They  are
well consistent with the theoretical predictions
\citep{Marietta2000,Meng2007}. Pakmor et al. (2008) simulated the
impact of type Ia SN on main sequence binary companions. They find
that the kick velocity of the companion star after the impact of the
SN ejecta varies from 17 to 61 km s$^{-1}$ for different models. Our
results are also consistent with their simulations.

We may also check the evolutionary state of the companion, assuming
that it filled the Roche lobe when the SN took place. Because the
Roche lobe radius is \citep{Paczynski1971}
$r_{r}=[0.38+0.2\log(\frac{1+\varepsilon}{1.4})]\times[\frac{1}{2.4+\varepsilon}(2.7\pm2.0)]\times10^{7}\approx(4.0\pm2.7)\times10^{6}$
km, about 5.7 times the solar radius, the companion should be a
subgiant, fully consistent with the luminosity classification of
Tycho G \cite{Ruiz2004} and \cite{Hern2009}.

The angular size (about 20$\degr$) of the
X-ray arc relative to the SN site is smaller
than that subtended by the Roche lobe (40$\degr$), which may be
expected from the compression and stripping of the envelope in the
bow-shocked SN ejecta material (Fig. 4).
 The angular separation
between the SN site and the arc is about half of the outer radius of
the remnant, which presumably reflects the difference in their
velocities. A type Ia SN explosion releases a typical kinetic energy
of (1-1.4)$\times10^{51}$ erg, and the mean ejecta velocity
($v_{ej}$) for a Chandrasekhar mass of 1.4 $M\odot$ is 8500-10000 km
s$^{-1}$. The momentum conservation for the stripped mass
($M_{str}$), the kicked companion star, and the ejecta in the solid
angle subtended by the envelope can be expressed as:
$M_{ej}v_{ej}$=$(M_{ej}+M_{str})\frac{v_{ej}}{2}+M_{c}v_k$.
Assuming a spherically symmetric SN and using a companion star mass
of 1 solar, an ejecta velocity of 9230 km s$^{-1}$, and $v_k$=42 km
s$^{-1}$,as well as the 20$\degr$ wide solid angle, we estimate the
mass of the envelope to be $0.0016$ $M\odot$. Taking the uncertainty
of $v_k$ ($\pm30$ km s$^{-1}$) into account, the upper limit of the
stripped mass is 0.0083 $M_{\odot}$. The mass outside the assumed solid
angle (e.g., to account for the full Roche lobe) should be
negligible because of the expected highly-concentrated radial mass
profile of a subgiant star \citep{Meng2007}. This estimation of the
stripped mass is consistent with that observed for two extragalactic
Ia SNe (Leonard 2007), significantly lower than the theoretical
predictions by Marrieta et al. (2000) and Meng et al. (2007), and is
close to (though still lower than) 1 to several percent that simulated 
by Pakmor et al. (2008).

Although the stripped mass is small, it is enough to produce the
X-ray arc. Katsuda et al. (2010) estimated
that the mean ambient density of Tycho's SNR is 0.0015 cm$^{-3}$,
or $<$0.2 cm$^{-3}$. The mass of the ambient ISM  in a cone of 20$\degr$  
radius should then be about
4$\times$10$^{-5}$ $M_{\sun}$, or $<$ 5$\times$10$^{-3}$ $M_{\sun}$,
smaller than or at most comparable to the stripped mass. Since bright
nonthermal filaments have been observed all along the rim
of the remnant (e.g., Cassam-Chenai et al. 2007), the
interaction between SN ejecta (probably denser at the
arc position) and the stripped mass should be strong
enough to be responsible of the X-ray arc emission.

Recently, Kerzendorf et al. (2009) have reported the new space
velocity and mass measurements of Tycho G, which are different from those by
Ruiz-Lapuente et al. (2004a). We list the new measurements and the
corresponding binary parameters in Table 2. The new orbital radius
and period are several times as those derived from
\cite{Ruiz2004} and \cite{Hern2009}. The angle between the kick and the 
space velocity of the stellar remnant is therefore about 82 degrees,
only marginally consistent with the numerical simulations (Marietta et al. 2000).
 In the scheme of momentum
conservation, the amount of the stripped mass is derived as
0.0087$\pm0.0017$ $M_{\sun}$. It is higher than that from \cite{Ruiz2004} and
\cite{Hern2009}, but still consistent with the simulations by
Pakmor et al. (2008) and the observations by Leonard (2007).

As pointed out by Kerzendorf et al. (2009), there is a simple
relationship between the companion's rotation velocity ($v_{rot}$)
and its orbital velocity ($v_{orb, 2})$:
$v_{rot}=\frac{M_1+M_2}{M_1}f(q)V_{orb,2}$, where $f(q)$ is the
ratio of the companion's Roche-lobe radius to the orbital separation
and $q=M_1/M_2$ is the mass ratio of the primary to the companion at
the time of the explosion. If $M_2$ is 0.3-0.5
$M_{\sun}$ (Kerzendorf et al. 2009), $v_{rot}$ of the companion's surface
should be about 24 km s$^{-1}$, much higher than 7.5 km s$^{-1}$,
the upper limit of Tycho G's rotation velocity
($v_{rot}\rm{sin}\it{i}$), where $i$ is the inclination angle of the
the orbital velocity. Using the inclination angle that we obtained in
this paper, the upper limit of $v_{rot}$ for Tycho G is about 10 km
s$^{-1}$, still significantly lower than 24 km s$^{-1}$.
Kerzendorf et al. (2009) proposed that a red giant scenario where the envelope's
bloating has significantly decreased the rotation could be consistent with
their observation of the low rotation velocity. Since the effect of the inclination
angle is small, Tycho G remains a stripped giant if it is the mass donor,
as suggested by Kerzendorf et al. (2009).

If the companion is about 1 $M_{\sun}$, its surface velocity was about 60 km s$^{-1}$ at
the SN explosion, 
as derived from the binary parameters listed in Table 2 and that it filled the 
Roche lobe, which has a radius of about 5.7 $R_{\sun}$. However, the companion 
may have a radius of 1-3 $R_{\sun}$ now \citep{Ruiz2004,Hern2009}, and so the 
surface velocity of the stellar remnant should be about 10-31 km s$^{-1}$, marginally
consistent with the upper limit of  $v_{rot}$ observed by \cite{kerz2009}
and the inclination angle that we obtained. We speculate 
that the decrease of the radius is possibly due to the destruction of the 
white dwarf. Before the SN explosion, the strong radiation of the accreting 
white dwarf inflated the envelope of the companion star to fill the Roche-lobe, and
the radius of the companion star shrinks to 1-3 $R_{\sun}$ now because the heating
of the white dwarf does not exist. In addition, we note here that
the shrink did not accelerate the rotation significantly, because the inflated 
envelope only contributes a small fraction of the total mass of the 
companion star and most of it was stripped away by the SN.

Using the position of the arc and the age of the remnant, we
obtained a mean velocity of the arc outward from the
SN site as about 0$\farcs$28 yr$^{-1}$. As measured by Katsuda et al. (2010),
the Tycho's remnant is expanding at a proper motion velocity around 0$\farcs$3 yr$^{-1}$,
in contrast to the mean expansion speed of about 0$\farcs$55 yr$^{-1}$ from
the radius and age of the remnant. This shows that the remnant are in a deceleration 
phase due to the interaction with the ISM. If the arc is a projected feature that is in the
outer layer of the remnant, it should have a proper motion
velocity of $\sim0\farcs$15 yr$^{-1}$. If the arc is actually about half way
from the SN site and represents the interaction of the SN explosion and the 
stripped companion envelope, the proper motion should be 
quite close to the mean velocity  $\sim$0$\farcs$28 yr$^{-1}$, as discussed below. 

On one hand, the arc is unlikely in an accelerating phase. 
The binary separation is tiny compared to the distance of the X-ray arc from the SN site. 
Even the Fe ejecta, which is expected to have the lowest velocity and was 
measured as $\sim3000$ km s$^{-1}$ currently (Furuzawa et al. 2009), can pass such a separation
within several hours. Most of the impact of the ejecta on the companion envelope (and thus 
the acceleration of the stripped mass) should take place soon after the explosion. On the 
other hand, the arc can not be significantly decelerated. Since the distance of the X-ray 
arc from the SN site is about half of the remnant  radius, using the ISM density given by 
Katsuda et al. (2010), the ISM mass in the cone  between the SN site and the X-ray arc 
is $\sim$5$\times$10$^{-6}$ $M_{\sun}$, much smaller than the stripped envelope mass.
Although the ISM decelerates the motion of the stripped envelope 
material, it is quite insignificant given the small mass. Therefore, the stripped envelope 
went through a very short accelerating phase in the beginning of the SN explosion, 
and has remained in an almost free expansion state since then.

We have attempted to measure the proper motion of the X-ray
arc so as to check the above scenario, since the expected proper
motion may be detectable at {\sl Chandra}'s resolution.
Unfortunately, the arc typically fallen more or less at a gap between
two ACIS-I chips, especially in
three  early observations. In addition, the arc is heavily
contaminated by strong thermal emission in the low energy band. In
4-6 keV, where nonthermal emission dominates, the
counting statistics of the arc are typically not sufficient in early
observations. As a result, we cannot yet get a reliable multi-epoch
measurements of the arc positions to allow for a reliable
determination of the proper motion. Future observations with more
careful positioning of the arc in the detector and with a total exposure time
comparable to the ACIS-I observations in 2009 will make such
measurements feasible.

\section{Summary}

We have shown that a self-consistent single-degenerate binary model
provides a natural and unified interpretation of the observed unique
X-ray arc/shadow in the Tycho's SNR. Two sets of parameters of the progenitor
binary system have also been presented using the optical observation results
of the candidate companion star (Tycho G) obtained by \cite{Ruiz2004} and \cite{kerz2009}
respectively. The main points in favor of our interpretation are:
(1) Although the nonthermal X-ray arc is half way from the remnant center, the high 
brightness show that it is viewed almost edge on and so unlikely a projected feature in
the outer layer of the remnant. Together with its sharp inward convex shape, the
arc most probably represents the interaction between the ejecta and a bulk of materials 
in the interior of the remnant.
(2) This bulk of materials can not be due to a pre-existing molecular cloud
or materials ejected by the progenitor binary system. The impact generating the X-ray
arc is most likely between SN ejecta and the stripped envelop of the companion star.
(3) The X-ray emission of the
remnant shows an apparent shadow casted by the arc in the
opposite direction of the explosion site, and there are local enhancements
in the same direction immediately within the X-ray arc, consistent with the
blocking of the SN ejecta by the envelope.
(4) We obtained a stripped mass of $\leq$ 0.0083 $M_{\sun}$, which is consistent with that
observed for two extragalactic Ia SNe (Leonard 2007) and close to the recent simulations by
Pakmor et al. (2008).
(5) The angle between the motion of the companion candidate and the direction of the arc as well
as the derived kick velocity of the companion star are well consistent with the theoretical
predictions and the numerical simulation results.

However, we note that there are still several points that can not be well interpreted by the 
current scenario, and further work is needed to reveal the physical processes related to
the nonthermal X-ray arc. (1) The properties of Tycho G and whether it is the stellar
remnant of Tycho's SN are under debate \citep{Ruiz2004,Hern2009,kerz2009}. If Tycho G is not
the stellar remnant, the binary parameters and kick velocity obtained in this paper
 are unreliable anymore. Also, if Tycho G is 
the stellar remnant but has properties as obtained by \cite{kerz2009}, the obtained angle
between the kick and the space velocity of the stellar remnant is only marginally consistent
with the numerical simulations \citep{Marietta2000}. (2) It has not 
been quantitatively estimated whether the interaction between the stripped envelope and the 
ejecta can produce shock wave strong enough to produce the nonthermal X-ray arc. 
(3) We failed to obtain a precise proper motion of the X-ray arc, which is an important criterion
to differentiate the X-ray arc as inside the remnant from a projected
feature in the outer layer.  Further studies of the stellar remnant as well as the measurement of the proper 
motion velocity of the X-ray arc are therefore urged to check the validity of the
scenario proposed in this paper.

\section*{Acknowledgments}
We thank the referee for many insightful comments and suggestions
that helped us greatly in the revision of the paper. This work is
supported by National Basic Research Program of China (973 program,
2009CB824800) and by National Science Foundation of China (10533020
and 10903007). QDW acknowledges the support by CXC/NASA under the
grant GO8-9047A and NNX10AE85G.

\clearpage

\begin{deluxetable}{ccc}
\tabletypesize{\footnotesize}
\tablecaption{The Chandra/ACIS-I observations of the Tycho SNR \label{tbl-1}}
\tablewidth{0pt}
\tablehead{
\colhead{ObsID}   & \colhead{Start Date}   &
\colhead{Exposure (ks)}}
\startdata
3837 & 2003-04-29 & 145  \\
7639 & 2007-04-23 & 109  \\
8551 & 2007-04-26 &  33\\
10093& 2009-04-13 & 118  \\
10094& 2009-04-18 & 90  \\
10095& 2009-04-23 & 173  \\
10096& 2009-04-27 & 106  \\
10097& 2009-04-11 & 107  \\
10902& 2009-04-15 & 40  \\
10903& 2009-04-17 & 24 \\
10904& 2009-04-13 & 35 \\
10906& 2009-05-03 & 41 \\
\enddata

%% Text for table notes should follow after the \enddata but before
%% the \end{deluxetable}. Make sure there is at least one \tablenotemark
%% in the table for each \tablenotetext.

%\tablenotetext{a}{Column density in unit of $10^{21}$cm$^{-2}$. All the components are forced to
%share the same column density}

%\tablecomments{The listed uncertainties are at the 90$\%$ confidence level. The fitting
%$\chi^2$ for all the five spectra are 366.4 with 386 degree of freedom. }
%Occasionally, authors wish to append a short
%paragraph of explanatory notes that pertain to the entire table, but
%which are different than the caption.  Such notes should be placed in
%a {\tt tablecomments} command like this.}

\end{deluxetable}

\clearpage

\begin{deluxetable}{lcc}
\tabletypesize{\footnotesize}
\tablecaption{The measurements of Tycho G and the binary parameters of Tycho SN's progenitor \label{tbl-2}}
\tablewidth{0pt}
\tablehead{
\colhead{Measurements} & \colhead{Ruiz-Lapuente04}   &
\colhead{Kernzendorf09}}
\startdata
Proper motion (mas yr$^{-1}$)
&$\mu_l$=-2.6$\pm1.34$& $\mu_l$=-1.6$\pm2.1$\\
 &$\mu_b$=-6.11$\pm1.34$ &$\mu_b$=-2.7$\pm1.6$\\
Tangential velocity (km/s)&94$\pm$41 & 51$\pm28$\\
 Radial velocity (km/s)
&50$\pm10$ & 49$\pm10$\\
Companion mass ($M\odot$) &1.0 & 0.3-0.5\\
 Orbital velocity (km/s) &98$\pm36$&71$\pm16$\\
 Orbital period (Day) & 4.9$^{+5.3}_{-3.0}$ & 28$\pm26$\\
 Separation (10$^{7}$ km) & 1.1$\pm0.7$& 3.2$\pm0.6$ \\
 Inclination angle ($\degr$) & 31$\pm13$ & 47$\pm20$\\
 Kick velocity (km/s) & 42$\pm30$ & 23$\pm20$\\
 Stripped Mass ($M\odot$) & 0.0016 ($<$0.0083) & 0.0087$\pm$0.0017\\
\enddata

%% Text for table notes should follow after the \enddata but before
%% the \end{deluxetable}. Make sure there is at least one \tablenotemark
%% in the table for each \tablenotetext.

\tablecomments{The velocity and mass values of Ruiz-Lapuente04 were
taken from Ruiz-Lapuente et al. (2004) and Hern\'andez et al. (2009),
and these of Kerzendorf09 were from Kerzendorf et al. (2009). The
uncertainties of all the adopted and derived parameters in this table are 
at 1$\sigma$ confidence level.}

\end{deluxetable}

\clearpage
\begin{figure}
\plotone{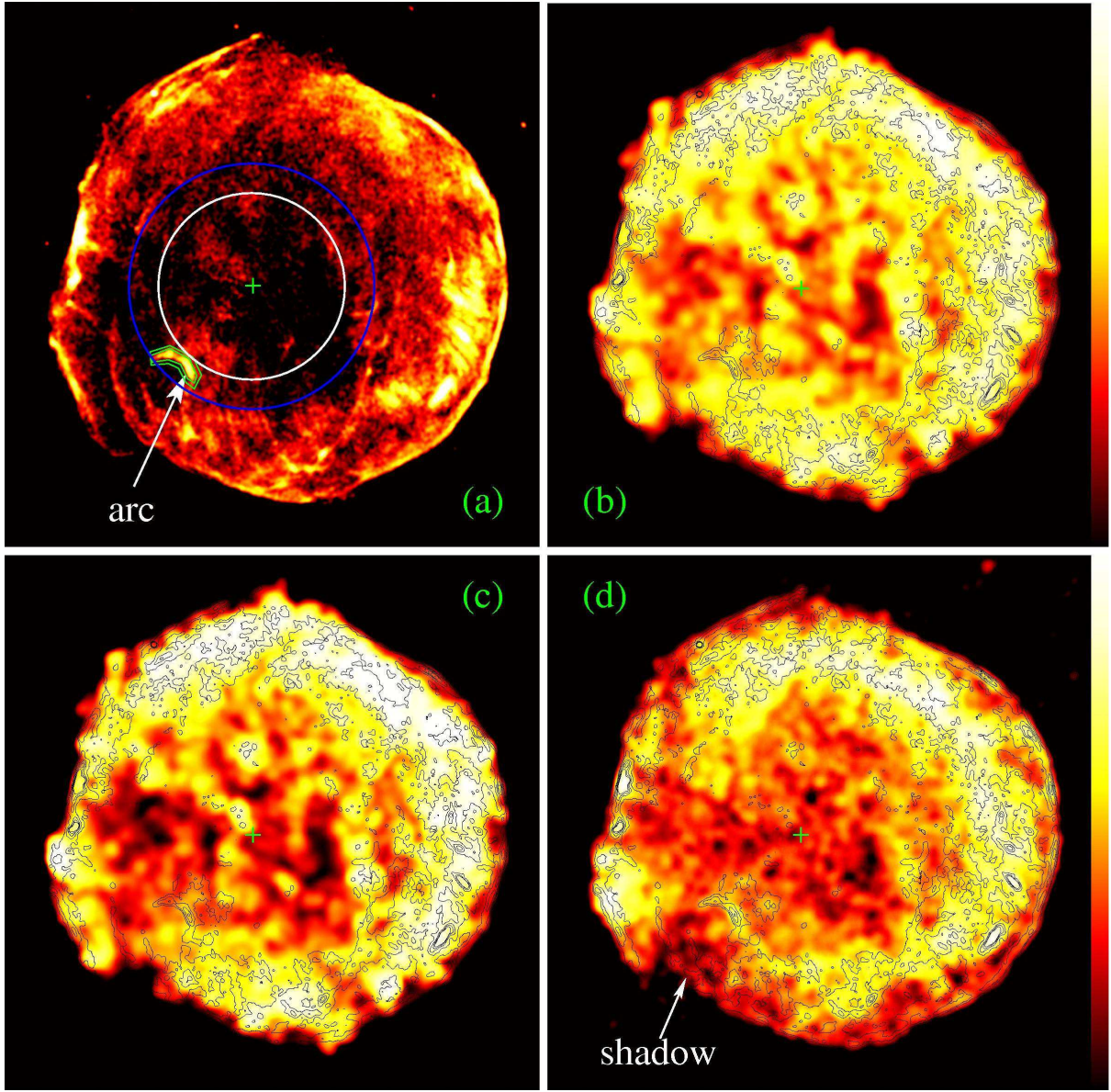}
\caption{ACIS-I intensity images of the Tycho SNR in the 4-6 keV (a), 1.6-2.0 keV (b), 2.2-2.6 keV (c) and
6.2-6.8 keV (d) bands, which represent the nonthermal continuum, Si, S, and Fe emission distributions respectively. The
images has been Gaussian-smoothed with the FWHM=3.5$\arcsec$ for (a) and 7.4$\arcsec$ for (b), (c), and (d).
The colour changes logarithmically from 7.4$\times$10$^{-6}$ to 1.5$\times$10$^{-4}$ counts cm$^{-2}$ s$^{-1}$ arcmin$^{-2}$ for (a),
1.5$\times$10$^{-4}$ to 5.9$\times$10$^{-3}$ counts cm$^{-2}$ s$^{-1}$ arcmin$^{-2}$ for (b),
 7.4$\times$10$^{-5}$ to 1.5$\times$10$^{-3}$ counts cm$^{-2}$ s$^{-1}$ arcmin$^{-2}$ for (c), and
2.2$\times$10$^{-6}$ to 3.0$\times$10$^{-5}$ counts cm$^{-2}$ s$^{-1}$ arcmin$^{-2}$ for (d). The contour levels represent the
4-6 keV emission and correspond
to 1.5$\times$10$^{-5}$, 3.0$\times$10$^{-5}$, 8.9$\times$10$^{-5}$, and 1.2$\times$10$^{-4}$ counts cm$^{-2}$ s$^{-1}$ arcmin$^{-2}$,
respectively. The images are produced from 12 ACIS-I observations listed in Table 1, the green crosses denote the
supernova explosion site inferred from the proper motion of Tycho G \citep{Ruiz2004}, the blue (white) circle defines
the inner (outer) boundary of the region from which the azimuth mean brightness profiles in the left (right) panel of 
Fig. 2  were produced.
\label{fig1}}
\end{figure}

\clearpage
\begin{figure}
\plotone{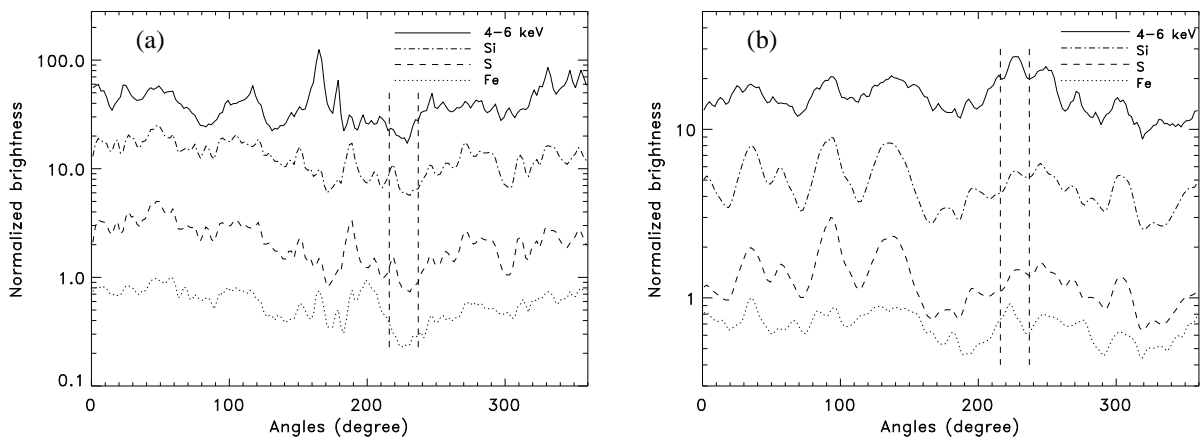}
\caption{(a): The azimuth mean brightness profiles of the Tycho SNR beyond the radius of the X-ray arc
in different energy bands as in Fig 1. Region used to produce the mean brightness profiles
is between the blue circle (Fig 1 (a)) and the outer boundary of the remnant. The maximum values
of the Fe, S, Si, and 4-6 keV curves were normalised to 1, 5, 25, and 125 so as to be displayed
more clearly. The angular range shadowed by the X-ray arc is between the two vertical dashed lines.
The angle is defined from west counter clock-wisely. (b): The same azimuth mean brightness profile of the
Tycho SNR as in the left panel but for regions ``inside'' the X-ray arc radius, i.e., within the white circle.
The maximum values of the Fe, S, Si, and 4-6 keV curves were normalised to 1, 3, 9, and 27 for the same reason
as in (a).
\label{fig2}}
\end{figure}

\clearpage

\begin{figure}
\plotone{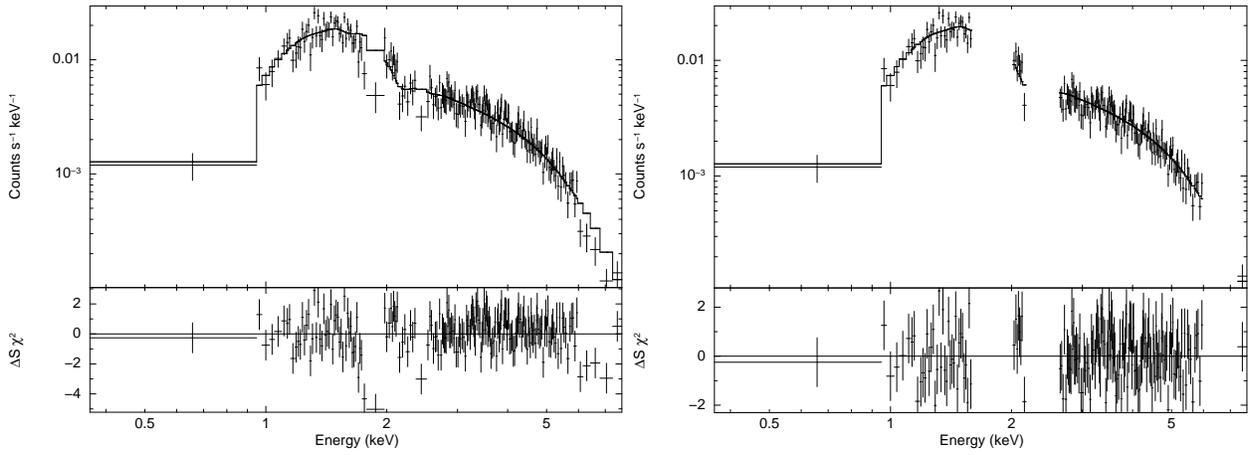}
\caption{Chandra ACIS spectrum of the X-ray arc, together with a power law model fit as described in the text.
In the spectral analysis, the background extracted from a region immediately surrounding the X-ray arc has
been subtracted.
\label{fig3}}
\end{figure}

\clearpage

\begin{figure}
\plotone{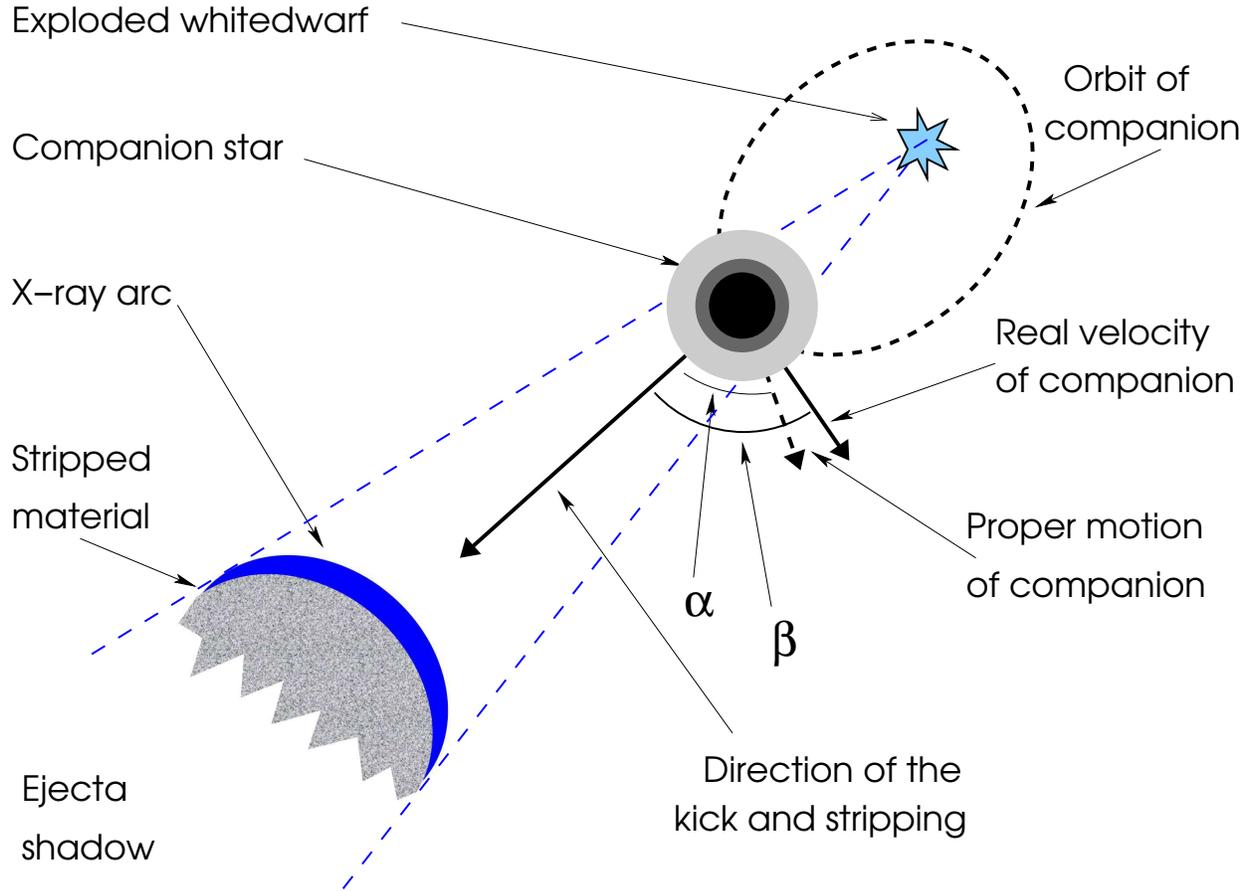}
\caption{Illustration of the progenitor binary and the impact of the SN on the companion star.
The companion star gained a kick in addition to the original orbital motion. Therefore, the
angle between the kick direction and the companion star velocity has a projected (observed)
value $\alpha= 63\degr$, but is $\beta$(= 67$\degr$) in real space. The outer envelop of the
companion star is stripped by the fast expanding SN ejecta in the same direction as the kick.
The observed X-ray arc represents the shock wave generated by the interaction between
the stripped envelope and the ejecta, which are behind the X-ray arc. The blocking of the
stripped envelope produces a cone that is deficient of SN ejecta.
\label{fig4}}
\end{figure}

\end{document}